\documentclass[conference]{IEEEtran}
\usepackage{graphicx}
\usepackage{color}
\usepackage{tabularx}
\usepackage{multirow}
\usepackage{comment}

\usepackage{cite}

\usepackage{amsmath}

\ifCLASSINFOpdf
\else
\fi
\usepackage{array}
\usepackage{amsmath}  
\usepackage{amssymb} 
\usepackage{amsthm} 
\usepackage{mathtools}

\newtheorem{theorem}{Theorem}

\def\qed{\hbox{${\vcenter{\vbox{
        \hrule height 0.4pt\hbox{\vrule width 0.4pt height 6pt
        \kern5pt\vrule width 0.4pt}\hrule height 0.4pt}}}$}}
 
\begin{document}

\title{The Logarithmic Random Bidding for the Parallel Roulette Wheel Selection with Precise Probabilities}

%




 \author{
\IEEEauthorblockN{Koji Nakano}
\IEEEauthorblockA{Graduate School of Advanced Science and Engineering, Hiroshima University\\
Kagamiyama 1-4-1, Higashihiroshima, 739-8527 Japan}
}


\newcommand{\h}{{\mathchar`-}}



\maketitle

\begin{abstract}
\boldmath
The roulette wheel selection is a critical process in heuristic algorithms, enabling the probabilistic choice of items based on assigned fitness values. It selects an item with a probability proportional to its fitness value. This technique is commonly employed in ant-colony algorithms to randomly determine the next city to visit when solving the traveling salesman problem. Our study focuses on parallel algorithms designed to select one of multiple processors, each associated with fitness values, using random wheel selection. We propose a novel approach called logarithmic random bidding, which achieves an expected runtime logarithmic to the number of processors with non-zero fitness values, using the CRCW-PRAM model with a shared memory of constant size. Notably, the logarithmic random bidding technique demonstrates efficient performance, particularly in scenarios where only a few processors are assigned non-zero fitness values.
\end{abstract}

\section{Introduction}
Let $f_0, f_1, \ldots, f_{n-1}$ denote non-negative real numbers referred to as \emph{fitness}.
\emph{The roulette wheel selection}, also known as \emph{the fitness proportionate selection},
is a fundamental operation that chooses one of the $n$ indices in a way that the probability of selecting an index $i$ ($0 \leq i \leq n-1$)
is directly proportional to its fitness value.
Mathematically, this probability of selecting $i$ is determined as:
\begin{align*}
F_i={f_i \over f_0+f_1+\cdots +f_{n-1}}.
\end{align*}
The roulette wheel selection serves as a crucial component in heuristic algorithms.
For instance, in the ant colony optimization technique for solving the traveling salesman problem (TSP)~\cite{Dorigo97, Pedemonte11, Uchida12},
this method is employed to choose the next city to visit.
Specifically, each edge connecting a pair of cities is assigned a fitness value based on its suitability as a TSP route.
Subsequently, the next city to visit is randomly selected from those connected to an unvisited city,
with the selection probability being directly proportional to the fitness value of the edge.
Furthermore, roulette wheel selection can be applied to the vertex coloring problem~\cite{Murooka16}.

This study focuses on parallel roulette wheel selection, aiming to efficiently select a processor $i$ among $n$ processors,
each assigned a fitness value $f_i$ corresponding to their ID from 0 to $n-1$.
The objective of parallel roulette wheel selection is to randomly select a processor $i$ with probability $F_i$.
For precise theoretical analysis, we assume the PRAM (parallel random access machine model)~\cite{Gibbons88},
equipped with multiple processors
and the shared memory with either EREW (exclusive read exclusive write) or CRCW (concurrent read concurrent write) operations.
All processors function synchronously, and in the EREW-PRAM model, simultaneous memory access by multiple processors is prohibited. 
On the other hand, in the CRCW-PRAM model, such simultaneous access is allowed.
In this CRCW-PRAM model, if a write conflict occurs in a memory cell of the shared memory, a randomly selected one among the multiple memory write operations succeeds in writing the value.

Utilizing a pseudo-random number generator rand() that returns a real number uniformly distributed in the range $[0,1)$,
we can implement the roulette wheel selection algorithm.
We define $p_i = f_0 + f_1 + \cdots + f_i$ ($0 \leq i \leq n-1$) as the prefix-sum, with $p_{-1} = 0$ for simplicity.
The parallel roulette wheel selection can be executed using the prefix-sum-based algorithm~\cite{Uchida12} outlined below:
\begin{tabbing}
xx \= xx\= xx\= \kill
[Prefix-sum-based parallel roulette wheel selection]\\
1. \> Compute all prefix-sums $p_0, p_1, \ldots, p_{n-1}$.\\
2. \> Processor 0 computes $R\leftarrow {\rm rand()}\cdot p_{n-1}$.\\
3. \> Each processor $i$ ($0 \leq i \leq n-1$) checks if $p_{i-1} \leq R < p_i$.\\
\>\> If this condition holds, processor $i$ is selected.
\end{tabbing}
This algorithm selects processor $i$ with probability:
\begin{align*}
{p_i-p_{i-1}\over p_{n-1}} &={f_i \over f_0+f_1+\cdots +f_{n-1}}=F_i.
\end{align*}
Consequently, the prefix-sum-based algorithm correctly selects $i$ with the probability $F_i$.
We can use a parallel prefix-sum algorithm on the EREW-PRAM~\cite{Gibbons88},
which operates in $O(\log n)$ time using $O(n)$ memory cells to compute all parallel prefix-sums.
Thus, the prefix-sum-based parallel roulette wheel selection runs in $O(\log n)$ time on the EREW-PRAM with a shared memory of size $O(n)$.

From a practical standpoint, the independent roulette selection~\cite{Cecilia13} might offer increased efficiency.
The algorithm for independent roulette wheel selection is outlined below:
\begin{tabbing}
xx \= xx\= xx\= \kill
[The independent roulette wheel selection]\\
1. \> Each processor $i$  ($0\leq i< n$)  computes $r_i= f_i\cdot rand()$.\\
2. \> Identify the maximum $r_i$ among $r_0, r_1, \ldots, r_{n-1}$\\
\>\> and select processor $i$.
\end{tabbing}
Since $r_i$ is a random number in the range $[0,f_i)$, a larger $f_i$ has a higher probability of becoming the maximum.
Thus, a processor with higher fitness is selected with a higher probability.
However, the probability may not equal to $F_i$.
This discrepancy can be observed through a simple example with $n=2$, $f_0=2$, and $f_1=1$.
In this case, $r_0 \in [0,2)$ and $r_1 \in [0,1)$.
If $r_0 \geq [1,2)$, then processor 0 is chosen.
Otherwise, since both $r_0$ and $r_1$ are in $[0,1)$, processor 0 is selected with a probability of $1 \over 2$.
Hence, processor 0 is chosen with a probability of ${1 \over 2} + {1 \over 2} \cdot {1 \over 2} = {3 \over 4}$ using the independent roulette wheel selection, while the roulette wheel selection requires selecting 0 with a probability of $F_0={2 \over {2+1}} = {2 \over 3}$.
Thus, the independent roulette wheel selection fails to adhere to the desired probabilities of the roulette wheel selection.
For a comprehensive analysis of selection probabilities in the independent roulette wheel selection, readers are encouraged to refer to~\cite{Lloyd17}.

This paper primary focuses on presenting a novel technique for the parallel roulette wheel selection method that ensures precise probabilities.
Our new technique called \emph{the logarithmic random bidding} is designed so that each $i$ ($0\leq i\leq n-1$)
is selected with a probability of $F_i$ within an expected time complexity of $O(\log k)$ on the CRCW-PRAM with a shared memory cell of size $O(1)$,
where $k$ denotes the number of non-zero fitness values among $f_0, f_1, \ldots, f_{n-1}$.
Hence, the logarithmic random bidding technique operates quite fast, especially when $k$ is small.
Notably, in ant-colony based TSP algorithms, fitness values are often set to zero for cities that have already been visited.
In such scenarios with many zero fitness values, the logarithmic random bidding technique exhibits accelerated performance.

This paper is organized as follows: Section~\ref{sec:logarithmic} introduces our logarithmic random bidding technique and demonstrates its ability to select processor $i$ with a probability of $F_i$.
Section~\ref{sec:implementation} details the implementation of a parallel roulette wheel selection with the logarithmic random bidding technique on the CRCW-PRAM.
Finally, Section~\ref{sec:concl} presents the conclusion of our work.

\section{The parallel roulette wheel selection with the logarithmic random bidding}
\label{sec:logarithmic}
This section presents \emph{the logarithmic random bidding} for the parallel roulette wheel selection and provides a proof ensuring that
it selects processor $i$ ($0 \leq i \leq n-1$) with a probability of $F_i$.
This algorithm closely resembles the independent roulette wheel selection, yet it incorporates a crucial modification in computing $r_i$.
Instead of the conventional $r_i = f_i \cdot \text{rand()}$ method used in the independent roulette wheel selection, it employs the logarithmic random bidding: $r_i = {\log({\rm rand()})\over f_i}$.
The specifics are detailed below:
\begin{tabbing}
xx \= xx\= xx\= \kill
[Roulette wheel selection with logarithmic random bidding]\\
1. \> Each processor $i$  ($0\leq i<n$)  computes $r_i={\log({\rm rand()})\over f_i}$.\\
2. \> Identify the maximum $r_i$ among $r_0, r_1, \ldots, r_{n-1}$\\
\>\> and select processor $i$.
\end{tabbing}

Next, we will demonstrate that the logarithmic random bidding technique selects processor $i$ ($0\leq i\leq n-1$)
with a probability of $F_i$.
It is evident that $r_i$ falls within the range $(-\infty, 0)$.
The cumulative distribution function of $r_i$ is
\begin{align*}
\Pr\left(r_i\leq x\right) &= \Pr\left({\log({\rm rand()})\over f_i}\leq x\right)\\ 
& =\Pr({\rm rand()}\leq e^{xf_i})\\
& = e^{xf_i}.
\end{align*}
Consequently, the probability density function of $r_i$ becomes:
\begin{align*}
{d\Pr(r_i \leq x) \over dx}&= f_i e^{xf_i}
\end{align*}
Let us evaluate the probability that processor 0 is selected, that is,
$r_0$ is larger than $r_1, r_2, \ldots, r_{n-1}$.
This probability can be evaluated as follows.
\begin{align*}
\MoveEqLeft \int_{-\infty}^0 {d\Pr(r_0 \leq x) \over dx}\cdot \Pr(r_1\leq x) \cdots \Pr(r_{n-1}\leq x) dx\\
&= \int_{-\infty}^0 f_0e^{xf_0} e^{xf_1} e^{xf_2} \cdots e^{xf_{n-1}}  dx \\
&=\int_{-\infty}^0  f_0e^{x(f_0+f_1+\cdots f_{n-1})} dx\\
&=\left[{f_0\over  f_0+f_1+\cdots +f_{n-1}}e^{x(f_0+f_1+\cdots f_{n-1})}\right]_{-\infty}^0 \\
&=  {f_0\over  f_0+f_1+\cdots +f_{n-1}}=F_0
\end{align*}
Therefore, processor 0 is selected with a probability of $F_0$.
Similarly, the probability of selecting any other processor $i$ can be determined in the same manner, confirming that processor $i$ is selected with a probability of $F_i$.

\begin{table}
\centering
\caption{The selection probabilities of the roulette wheel selection algorithms in $10^9$ iterations with $f_i=i$ ($0\leq i\leq 9$)}
\label{tab:probability}
\begin{tabular}{c|c|c|c|c}
$i$ & $f_i$ & $F_i$ &independent  & logarithmic  \\
\hline
0 &0	&0.000000 	&0.000000 	&0.000000 \\ 
1&1	&0.022222 	&0.000000 	&0.022222 \\
2&2	&0.044444 	&0.000088 	&0.044446 \\
3&3	&0.066667 	&0.001708 	&0.066672 \\
4&4	&0.088889 	&0.010993 	&0.088885 \\
5&5	&0.111111 	&0.038787 	&0.111105 \\
6&6	&0.133333 	&0.094267 	&0.133340 \\
7&7	&0.155556 	&0.178238 	&0.155552 \\
8&8	&0.177778 	&0.282382 	&0.177771 \\
9&9	&0.200000 	&0.393536 	&0.200007 
\end{tabular}
\end{table}

\begin{table}
\centering
\caption{The selection probabilities of the first 10 processors of the roulette wheel selection algorithms in $10^9$ iterations with $f_0=1$ and $f_1=f_2\cdots =f_{99}=2$}\label{tab:probability100}
\begin{tabular}{c|c|c|c|c}
$i$ & $f_i$ & $F_i$ &independent  & logarithmic  \\
\hline
0&	1	&0.005025	&0.000000 	&0.005026 \\
1&	2	&0.010050	&0.010104 	&0.010053 \\
2&	2	&0.010050	&0.010104 	&0.010053 \\
3&	2	&0.010050	&0.010099 	&0.010048 \\
4&	2	&0.010050	&0.010101 	&0.010050 \\
5&	2	&0.010050	&0.010108 	&0.010057 \\
6&	2	&0.010050	&0.010099 	&0.010048 \\
7&	2	&0.010050	&0.010101 	&0.010050 \\
8&	2	&0.010050	&0.010100 	&0.010050 \\
9&	2	&0.010050	&0.010106 	&0.010055 
\end{tabular}
\end{table}

To verify the probability precision of the logarithmic random bidding technique against the conventional independent roulette wheel selection, 
we conducted simulations over $10^9$ iterations using $f_i=i$ ($0 \leq i \leq 9$).
The Mersenne Twister random number generator~\cite{Matsumoto98} was employed to implement the rand() function.
The comparative analysis showcases the inaccuracy of the independent roulette wheel selection, particularly evident for smaller $f_i$ values where the selection probability significantly deviates from $F_i$.
In contrast, our roulette wheel selection consistently demonstrates probabilities of selecting $f_i$ that closely align with $F_i$.

Especially for large $n$, the probability of selecting smaller $f_i$ values can tend toward near-zero values.
For instance, in a scenario employing the roulette wheel selection across 100 processors with $f_0=1$ and $f_1=f_2=\cdots=f_{99}=2$, the expected selection probability for processor 0 stands at $\frac{1}{199} \approx 0.005025$.
However, the independent roulette wheel selection yields a probability of $({\frac{1}{2}})^{99} \cdot \frac{1}{100} \approx 1.57772 \times 10^{-32}$, essentially zero, resulting in processor 0 never being selected.
Table~\ref{tab:probability100} shows simulations from $10^9$ iterations, confirming alignment with mathematical analyses: the independent roulette wheel selection consistently neglects selection of processor 0, while the logarithmic random bidding technique accurately reflects the expected selection probabilities.

\section{Implementation of the parallel roulette wheel selection algorithm}
\label{sec:implementation}
This section focuses on implementations of the parallel roulette wheel selection.
In particular, we will present how we can identify the maximum $r_i$ among $r_0, r_1, \ldots, r_{n-1}$
in parallel.
It can be identified by a parallel reduction in an obvious way as follows.
Imagine a binary tree with $n$ leaves each associated with $r_i$.
The maximum of the two children are computed in very internal nodes from the leaves synchronously.
Clearly, the root will store the  maximum.
However, this implementation takes $O(\log n)$ time and requires a shared memory of size $O(n)$ on the EREW-PRAM.

We aim to demonstrate that the maximum value $r_i$ can be identified within an expected $O(\log k)$ parallel steps.
The parallel algorithm utilizes the CRCW-PRAM model, employing shared memory variable $s$ initialized to zero, along with the variable ${\it output}$
to store the index of selected processor.

Each processor $i$ iteratively writes its $r_i$ value into $s$ until the condition $s\geq r_i$ is satisfied.
Consequently, write conflicts may arise, with one of the writing accesses successfully updating the shared memory cell.
Once all processors complete this writing operation and the condition $s\geq r_i$ is met for all $i$, the value stored in $s$ represents the maximum of all $r_i$'s.
Each processor $i$ writes its index $i$ into ${\it output}$ if $s=r_i$.
Clearly, ${\it output}$ retains the index corresponding to the maximum $r_i$ value.
Here are the detailed steps:
\begin{tabbing}
xx \= xx\= xx\= \kill
[Identifying the maximum $r_i$]\\
Each processor $i$ ($0\leq i< n-1$)  performs:\\
1. \> while $s<r_i$ do $s\leftarrow r_i$;\\
2. \> barrier\_synchronization();\\
3. \> if $s=r_i$ then ${\it output} \leftarrow i$;
\end{tabbing}

Next, we will estimate the number of iteration performed in the while loop.
We assume that we have $k$ non-zero $r_i$'s.
We say that a processor $i$ is \emph{active} in an iteration of the while loop
if $s<r_i$ holds and it performs $s\leftarrow r_i$.
Clearly, the first iteration, $k$ processors are active and the while loop is iterated until
no active processor exists.
We say that the an iteration is success if at least a half of active processors
become inactive.
The iterations can have up to $\lceil\log_2 k\rceil$ success iterations
and every iteration is success with probability $1\over 2$ when $k\geq 2$.
The expected number of iterations is
\begin{align*}
1\cdot {1\over 2}+2\cdot{1\over 2^2}+ 3\cdot{1\over 2^3}+\cdots &= 2.
\end{align*}
Hence, $2\lceil\log_2 k\rceil$ iterations are sufficient to have $\lceil\log_2 k\rceil$ success iterations,
and we have the following theorem:
\begin{theorem}
The parallel roulette wheel selection with the logarithmic random bidding for $n$ processors, each with a non-negative fitness $f_i$ ($0\leq i\leq n-1$), selects a processor $i$ with a probability of $F_i={f_i \over f_0+f_1+\cdots +f_{n-1}}$ in $O(\log k)$ expected time on the CRCW-PRAM with a shared memory size of $O(1)$, where $k$ is the number of non-zero fitness values among all $f_i$'s.
\end{theorem}

\section{Conclusion}
\label{sec:concl}
This paper presented the logarithmic random bidding technique for the parallel random wheel selection with precise selection probabilities.
It runs in $O(\log k)$ time on the CRCW-PRAM with a shared memory of size $O(1)$ and the selection probabilities
follow the requirement of the roulette wheel selection precisely.

\bibliographystyle{IEEEtran}
\bibliography{algorithm,gpu,fpga,sort,mining,halftone}

\end{document}